\documentclass[aps,prl,twocolumn,groupedaddress,showpacs]{revtex4}

\usepackage{graphicx}

\begin{document}

\title{The giant plasticity of a quantum crystal}

\author{Ariel Haziot$^1$, Xavier Rojas$^1$, Andrew D. Fefferman$^1$,
John R. Beamish$^{1,2}$, and S\'ebastien Balibar$^1$}

\affiliation{1- Laboratoire de Physique Statistique de l'Ecole Normale
Sup\'erieure, associ\'e au CNRS et aux Universit\'es P.M. Curie and D.
Diderot, 24 rue Lhomond, 75231 Paris Cedex 05, France.\\
2 - Departement of Physics, University of Alberta, Edmonton, Alberta
Canada T6G 2G7}

\begin{abstract}
When submitted to large stresses at high temperature, usual crystals
may irreversibly deform.  This phenomenon is known as plasticity and
it is due to the motion of crystal defects such as dislocations.  We
have discovered that, in the absence of impurities and in the zero
temperature limit, helium 4 crystals present a giant plasticity that
is anisotropic and reversible.  Direct measurements on oriented single
crystals show that their resistance to shear nearly vanishes in one
particular direction because dislocations glide freely parallel to the
basal planes of the hexagonal structure.  This plasticity disappears
as soon as traces of helium 3 impurities bind to the dislocations or
if their motion is damped by collisions with thermal phonons.
\end{abstract}

\pacs{67.80.de, 67.80.dj, 62.20.fq, 62.20.de}
\maketitle

A crystal resists elastically to a shear stress, contrary to a fluid,
which flows.  When applied to a crystal, a small shear stress $\sigma$
produces a strain $\varepsilon$ that is proportional to $\sigma$.  If
the stress is released, the crystal returns elastically to its
original shape.  The elastic shear modulus of the crystal is the ratio
$\mu=\sigma/\varepsilon$.  However, if the applied stress is large, a
usual crystal is "plastic": it deforms more than in the elastic regime
thanks to the motion of dislocations.  With classical crystals,
plastic deformation is irreversible and small.  In metals, it requires
stresses of order 10$^{-1}$ to 10$^{-6}$ of the shear modulus and the
strain rate increases with temperature \cite{Vardanian,Tinder}.  We
have discovered that, in the case of ultrapure helium 4 single
crystals around 0.1~K, the resistance to shear nearly vanishes in one
particular direction while in another direction no measurable
deviation from normal elastic behavior occurs.  This giant
anisotropic plasticity is independent of stress down
to extremly low values (10$^{-11}$ times the elastic shear modulus).
We demonstrate that it is a consequence of dislocations gliding freely along the
basal planes of the hexagonal crystal structure. Some gliding
along these planes had been observed
\cite{Berghezan,Hull} in other hexagonal crystals but never with
comparable amplitude.  In $^4$He crystals, we show that dislocations
are able to move at high speed with negligible dissipation so that
the plasticity is also independent of the rate at which the strain is
applied.  Moreover,
this motion is reversible so that the plasticity shows up as a 50 to
80\% reduction of one particular elastic coefficient, which we have
identified as $c_{44}$.  The plasticity disappears if traces of
impurities bind to dislocations or if thermal phonons damp their
motion.  Our observations have been made possible by direct elasticity
measurements on oriented crystals where the impurity concentration
could be lowered from 0.3 ppm to zero at very low temperature (15 mK).
The gliding of dislocations is an intensively studied phenomenon of
fundamental importance in Materials Science \cite{Proville}. We
provide strong experimental evidence for its precise origin.

\begin{figure}[t]
\includegraphics[width=\linewidth]{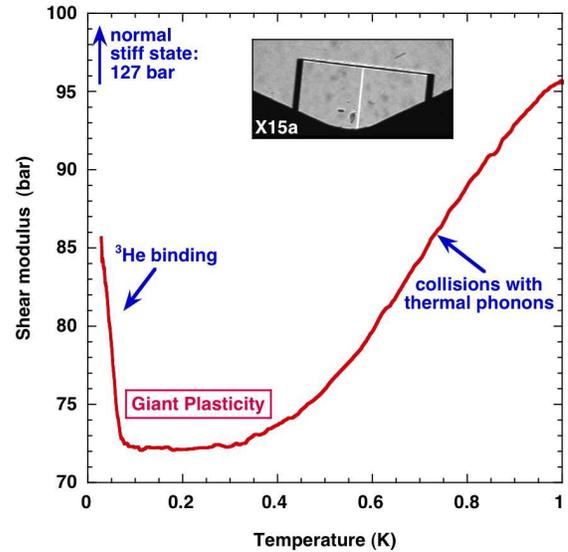}
\caption{The shear modulus of crystal X15a
as a function of temperature.
Around 0.2~K, the shear modulus is reduced to 72~bar, much less
than the normal value (127 bar).}
\label{Fig1-plasticity}
\end{figure}

We have improved some of our previous
techniques \cite{Day07,Sasaki08,Pantalei,Rojas}.  Here we briefly
describe them (see details in the Supplemental Material).  In a
transparent cell inside a dilution refrigerator with optical access,
crystals can be seen during growth.  After determining their
orientation from their shape (Fig.~1), we grow them in a 1~mm gap
between two piezo-electric transducers.  Applying a small ac-voltage
to one transducer (0.001 to 1 Volt, 200 to 20,000 Hz) produces a
vertical displacement (0.001 to 1 Angstrom) and consequently a strain
$\varepsilon$ of amplitude 10$^{-10}$ to 10$^{-7}$ and a stress
$\sigma=\mu \varepsilon$ (10$^{-9}$ to 10$^{-5}$ bar) on the other
transducer ($\mu$ is the relevant shear modulus of the crystal).  The
stress generates a current that is our signal after amplification with
a lock-in.  The shear modulus $\mu$ and the dissipation $1/Q$ are
obtained from the magnitude and phase of the signal.  Various growth
methods lead to three different types of crystals \cite{Sasaki08}: \\
1 - High quality "type 1" crystals are grown by pressurizing the
superfluid liquid in the cell at 20 mK up to the crystallization
pressure and staying there after the cell is nearly full.  Random
nucleation produces different orientations.  In this case, there
remains some liquid in corners but not in the gap.  Fresh from growth,
such crystals are totally free of impurities, especially if one uses
ultrapure $^4$He containing 0.4 ppb of $^3$He (we also use natural
$^4$He with 0.3 ppm $^3$He).  Slow growth at 20 mK rejects $^3$He
impurities in front of the liquid-solid interface as in the "zone
melting" purification of metals but here this method is extremely
efficient because the temperature is very low \cite{Pantalei}.
However, when warming such ultrapure crystals, $^3$He impurities come
back into the crystal where their concentration varies with
temperature.\\
2 - Single crystals of lower quality (type 2) are grown at 1.4 K. All
the remaining liquid is crystallized during further cooling along the
melting line from 26.1 bar at 1.4 K to 25.3 bar at 1 K, below which it
stays constant \cite{Balibar05}.  This pressure change probably creates
dislocations, but the $^3$He concentration is that of the helium gas
used to grow them.  Any type 1 crystal can be melted down to a small
seed and regrown at 1.4 K as a type 2 crystal with the same
orientation, or vice versa.\\
3 - Polycrystals (type 3) are grown at constant volume starting with
normal liquid at 3 K and 60 bar.
\begin{figure}
\includegraphics[width=\linewidth]{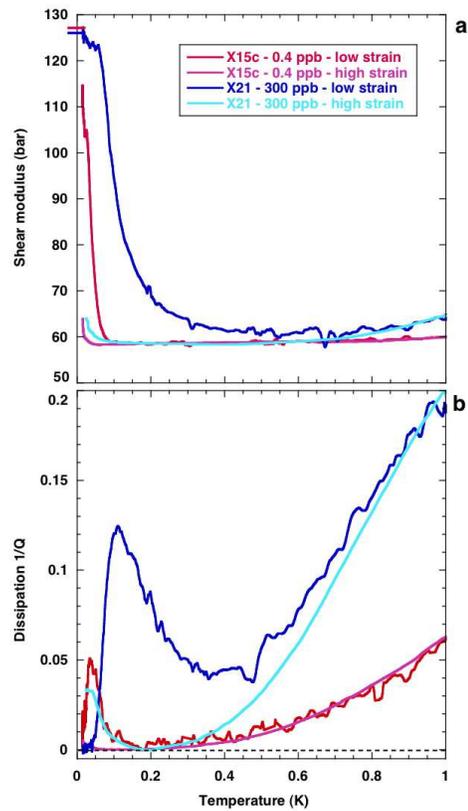}
\caption{Comparison of two "type 2" crystals: X15c and X21 grown
respectively from ultrapure and from natural
$^4$He.  The $T$-domain of the giant plasticity
depends on both purity and strain
amplitude.  On graph (a), two ticks on the vertical axis indicate the
predicted value of the shear modulus for immobile dislocations.  The
lower graph (b) shows that the dissipation is negligible in the high
plasticity region of X15c, as for immobile dislocations in the zero
temperature limit of X21.}
\label{Fig2-plasticity}
\end{figure}

Fig. 1 presents a measurement of the shear modulus $\mu$ of crystal
X15a at 9 kHz during cooling. This is an ultrapure type 1 crystal,
although grown at 600 mK. Its shear modulus reaches a very small value
around 0.2 K. At lower $T$, the crystal
stiffens as a consequence of dislocation pinning
by $^3$He \cite{Day07,Rojas,Syshchenko10}.  Crystals with
immobile dislocations are stiffer than with mobile dislocations. The
pinning temperature depends on the $^3$He concentration, on the
binding energy $E_3$ = 0.73 K \cite{Syshchenko10}, and on the
measurement frequency.  Above 0.3 K, $\mu$ also increases, probably
because dislocations scatter thermal phonons \cite{Iwasa79}.
We thus understand that $^4$He crystals show an
anomalous shear modulus between impurity pinning at very low $T$ and
collisions with thermal phonons at higher $T$. We call it a "giant plasticity"
because it is
due to the large motion of dislocations under very small stresses.
Since it is reversible, the plasticity
shows up as a reduction in the shear modulus.  Note that the present
measurements are made at kHz frequencies during hours with results that are
independent of the sign and duration of the applied strain.

Two type 2 crystals are compared in Fig.~2a: X15c (purity 0.4 ppb)
and X21 (0.3 ppm $^3$He), at high or low driving strain.  When a
larger drive is applied, the temperature at which $^3$He bind to
dislocations is lower, so that the plasticity domain is
wider.  As expected \cite{Day07,Rojas,Syshchenko10},
pinning by $^3$He occurs at lower $T$
for ultrapure crystals.  Fig. 2b shows the corresponding dissipation.
In the plasticity domain of X15c, we
measure zero dissipation, as in the low $T$ limit of X21 where $^3$He
binding suppresses dislocation motion.  The peak
dissipation corresponding to $^3$He binding is lower in temperature
for the ultrapure crystal X15c and it nearly disappears at a high
strain $\varepsilon=10^{-7}$ that exceeds the threshold for $^3$He
unbinding (3 10$^{-8}$, see \cite{Day07,Day12}). The dissipation
above 0.3 K is independent of amplitude and increases with $T$ as expected
for a thermal phonon scattering mechanism \cite{Iwasa79}.

\begin{figure}[t]
\includegraphics[width=\linewidth]{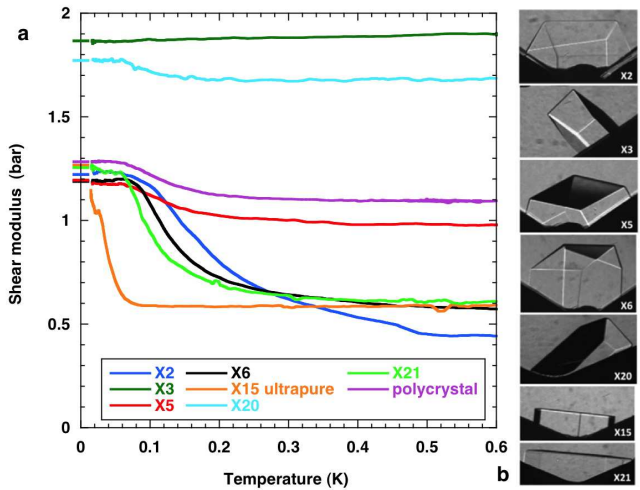}
\caption{Measurements (a) for different crystal orientations show that
the plasticity is highly anisotropic.
Ticks on the vertical axis show the predicted shear
modulus with immobile dislocations, in excellent agreement with
measurements in the low temperature limit where dislocations are
pinned by $^3$He.}
\label{Fig3-plasticity.eps}
\end{figure}

We have identified the precise origin of the plasticity with the
following quantitative study.  Hexagonal crystals have 5 independent
elastic coefficients \cite{Day12,Crepeau,Greywall} $c_{11}$, $c_{12}$,
$c_{13}$, $c_{33}$, $c_{44}$ and $c_{66}$ = ($c_{11}$-$c_{12}$)/2.
The indices 1 to 6 respectively refer to xx, yy, zz, yz, xz and xy,
where z is the [0001] "c" axis.  Fig.~3 shows results from
7 type 2 crystals, all of natural purity except the ultrapure X15,
and one polycrystal (BC2). X3 is tilted
with a c-axis 45$^\circ$ from the vertical. Its effective shear
modulus is $\mu$ = 0.004 $c_{44}$ + 0.004 $c_{66}$ +
0.25 ($c_{11}$+$c_{33}$-2$c_{13}$), practically independent of
$c_{44}$ and $c_{66}$ (see Suppl. Mat.). It shows no
$T$-dependence.  Since it is very unlikely that these coefficients
would vary with $T$ in just such a way to
keep ($c_{11}$+$c_{33}$-2$c_{13}$) constant, we conclude
that $c_{12}$, $c_{13}$ and $c_{33}$ are
independent of $T$, so that X3 could be used to
calibrate the piezo-electric coefficient of our transducers
(0.88 Angstrom/Volt in our first cell, 0.95 in the second
cell).  For this we used values of elastic coefficients measured
from ultrasound velocity at the melting pressure
\cite{Crepeau,Greywall} $c_{11}$ = 405 bar, $c_{12}$ = 213 bar,
$c_{13}$ = 105 bar, $c_{33}$ = 554 bar, $c_{44}$ = 124 bar, and
$c_{66}$ = 96 bar.  These values correspond to the intrinsic
elasticity of $^4$He crystals because dislocations could not move
at 10~MHz above 1.2~K. Given this calibration we could
calculate the shear modulus for all other orientations in the $T$=0
limit (ticks on the vertical axis of Fig. 3).
Excellent agreement is found with our measurements except for
X15 whose stiffening was not completed at 15 mK. For the
polycrystal BC2, Maris' averaging method \cite{Maris10}
leads to excellent agreement again.

All crystals except X3 show a strong $T$-dependence, which could be
due to a reduction of either $c_{44}$ or $c_{66}$.  Indeed, hexagonal
crystals usually have one preferential gliding plane that can be
either the basal plane (0001) or the prismatic planes perpendicular
to it.  For hcp crystals, it is predicted to be
the basal plane, although directional bonding in hexagonal metals can
lead to non-close packed structures which favour prismatic glide
\cite{Berghezan,Hull}. For X21, the dependence on $c_{66}$
is negligible and the large variation of $c_{44}$ could be easily
extracted.  Assuming then that $c_{66}$ is constant, we could
calculate the variation of c$_{44}$ in all crystals.  Since
X2, X5, X6 and X21 were all type 2 crystals grown in
similar conditions, one expects the $T$-dependence of their
coefficient c$_{44}$ to be the same. Fig.~4
shows that this assumption leads to a reproducible reduction of
$c_{44}$ (62$\pm$8 \% for all type 2 crystals), although the shear
modulus of X2 depends mostly on $c_{44}$ while that of X5 depends
mostly on $c_{66}$.  On the opposite, assuming that $c_{44}$ is
constant and $c_{66}$ varies leads to absurd results: from 60\%
reduction in $c_{66}$ for X5 to 300\% for X6 and to 1000 \% for X2,
not mentioning X21.  The preferential direction for gliding has to be
the basal plane. Only $c_{44}$ depends on temperature. This is why
Fig.~3 shows a highly anisotropic plasticity.

\begin{figure}
\includegraphics[width=\linewidth]{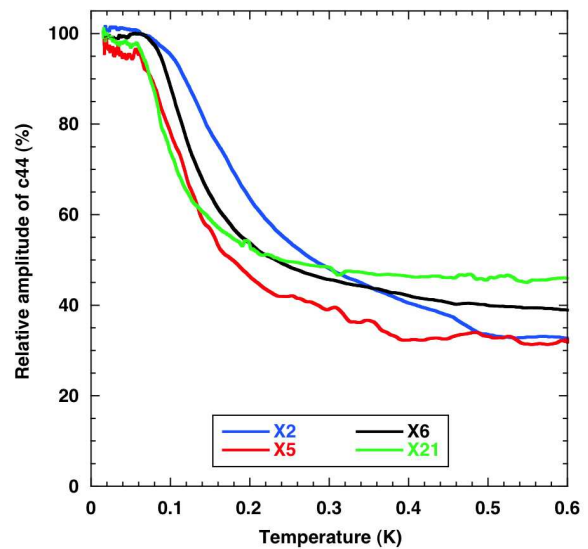}
\caption{Assuming that dislocations glide in the basal plane,
one finds the same variation of $c_{44}$ by 62$\pm$8 \% for X2, X5, X6 and X21, which are crystals with different
orientations grown in the same conditions. Assuming that they glide
in the prismatic planes leads to absurd results (see text).}
\label{Fig4-plasticity.eps}
\end{figure}

The amplitude dependence is also consistent with $^3$He pinning at low
$T$. Fig. 5 shows the shear modulus as a function of stress for 4
crystals at 20 mK. Above a 1 $\mu$bar threshold due to the breakaway from
$^3$He impurities \cite{Day07,Day12}, the
shear modulus decreases. By projecting the stress on the
basal plane, we determined the "resolved stress" (see Suppl.
Mat.), and we obtained a reproducible threshold for different
orientations, confirming that dislocations glide in the basal planes.  This
threshold is hysteretic because the force acting on dislocations
depends on the pinning length $L_i$ between
impurities \cite{Iwasa80}.  One now understands why Fig.~2 shows a
$^3$He pinning occuring at lower $T$ with a large strain (10$^{-7}$)
than with a low strain (10$^{-9}$).  In contrast, the phonon damping
is independent of the strain amplitude.

\begin{figure}
\includegraphics[width=\linewidth]{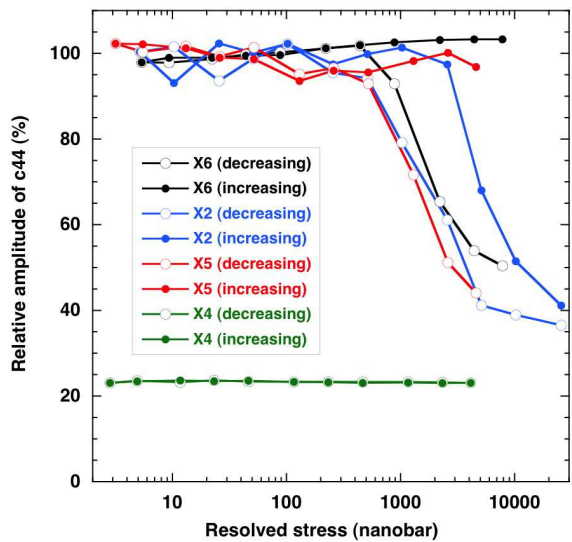}
\caption{The relative amplitude of $c_{44}$ for 4 different crystals
at 20 mK as a function of the stress projected on the basal plane
(''resolved'').  At a threshold stress around 1 microbar,
dislocations break away from $^3$He impurities\cite{Day07,Day12}.
This threshold is larger when increasing the stress than when
decreasing it.  Being free of impurities, X4 shows a reduction of
$c_{44}$ by 80\%, independent of stress.}
\label{Fig5-plasticity.eps}
\end{figure}

Fig.~5 shows additional measurements made on X4, a type 1 crystal that
was cooled down under high strain (10$^{-7}$) at 3 kHz in order to
expell all $^3$He impurities into the liquid.  Its coefficient
$c_{44}$ is reduced by 80\% even at 20 mK. This value is stable in
time ($^3$He stay trapped in the liquid), and independent of the
strain amplitude.  After \textit{assuming} that dislocations glide in
the basal plane, Rojas \textit{et al.} \cite{Rojas} found 86\%
reduction for a type 1 crystal of probably better quality (see
Suppl. Mat.).  Here we \textit{demonstrate} that this assumption was
correct, and that the plasticity is very large in particular
directions.  We have not yet prepared a crystal with
zero dislocation as might be possible \cite{Ruutu}.
Dislocations arranged in a simple Frank
network should lead to less than 5\% reduction in $c_{44}$
\cite{Friedel}.  In our case, they may be aligned in sub-boundaries
\cite{Iwasa95} where they cooperate to reduce $c_{44}$ much more.
Assuming a typical dislocation density \cite{Syshchenko08} $\Lambda$
=10$^4$ cm$^{-2}$, we could estimate the amplitude of their motion.
With a 10~kHz strain $\varepsilon$ = 3 10$^{-8}$, we find that dislocations
move a distance $d \sim \varepsilon/\Lambda b$ = 1 $\mu$m at
1 cm/s ($b$ = 3 Angstrom is the Burgers vector
amplitude).  This is 30 million Burgers vectors per second, a giant
effect with no equivalent in classical crystals
(see Suppl. Mat.)\cite{Hikata,Farla,Friedel}.

In summary, we have demonstrated that the plasticity of $^4$He
crystals is reversible and anisotropic, due to the free gliding of
dislocations parallel to the basal planes.  Classical plasticity is
irreversible because dislocations jump from one pinning site to
another and multiply, leading to "work hardening".  In our case,
applying large stresses at low $T$ produces defects, which harden the
crystals and are probably jogs because annealing at 1 K restores the
high crystal quality (see Suppl. Mat.).  For simplicity, we
show measurements in this article, which are recorded on cooling after
annealing.  The giant plasticity of $^4$He crystals is mainly due to
their absolute purity at very low temperature.  An interesting
question is whether dislocations move by tunnelling or by thermal
activation over Peierls barriers.  Tunnelling is suggested by the
reversible motion at low $T$ without dissipation or strain-dependence.
However, quantum fluctuations probably reduce the kink energy E$_k$ by
enlarging their width.  If $k_BT > E_k$, kinks are not relevant, as is
the case for surface steps \cite{Nozieres,Rolley} whose dynamics is
also dissipation free in the low $T$ limit.  A calculation of E$_k$
could indicate if the dislocation motion is quantum or classical
thanks to a negligible kink energy.

This work was supported by grants from ERC (AdG
247258) and from NSERC Canada.

\end{document}


\title{The giant plasticity of a quantum crystal - Supplemental Material}

\author{Ariel Haziot$^1$, Xavier Rojas$^1$, Andrew D. Fefferman$^1$,
John R. Beamish$^{1,2}$, and S\'ebastien Balibar$^1$}

\affiliation{
1- Laboratoire de Physique Statistique de lÕEcole Normale
SupŽrieure, associŽ au CNRS et aux UniversitŽs P.M. Curie and D.
Diderot, 24 rue Lhomond, 75231 Paris Cedex 05, France.\\
2 - Departement of Physics, University of Alberta, Edmonton, Alberta
Canada T6G 2G7}

\pacs{67.80.bd, 67.80.de, 67.80.dj}

\maketitle

\section{The experimental cell and the transducer calibration}

\begin{figure}
\includegraphics[width=\linewidth]{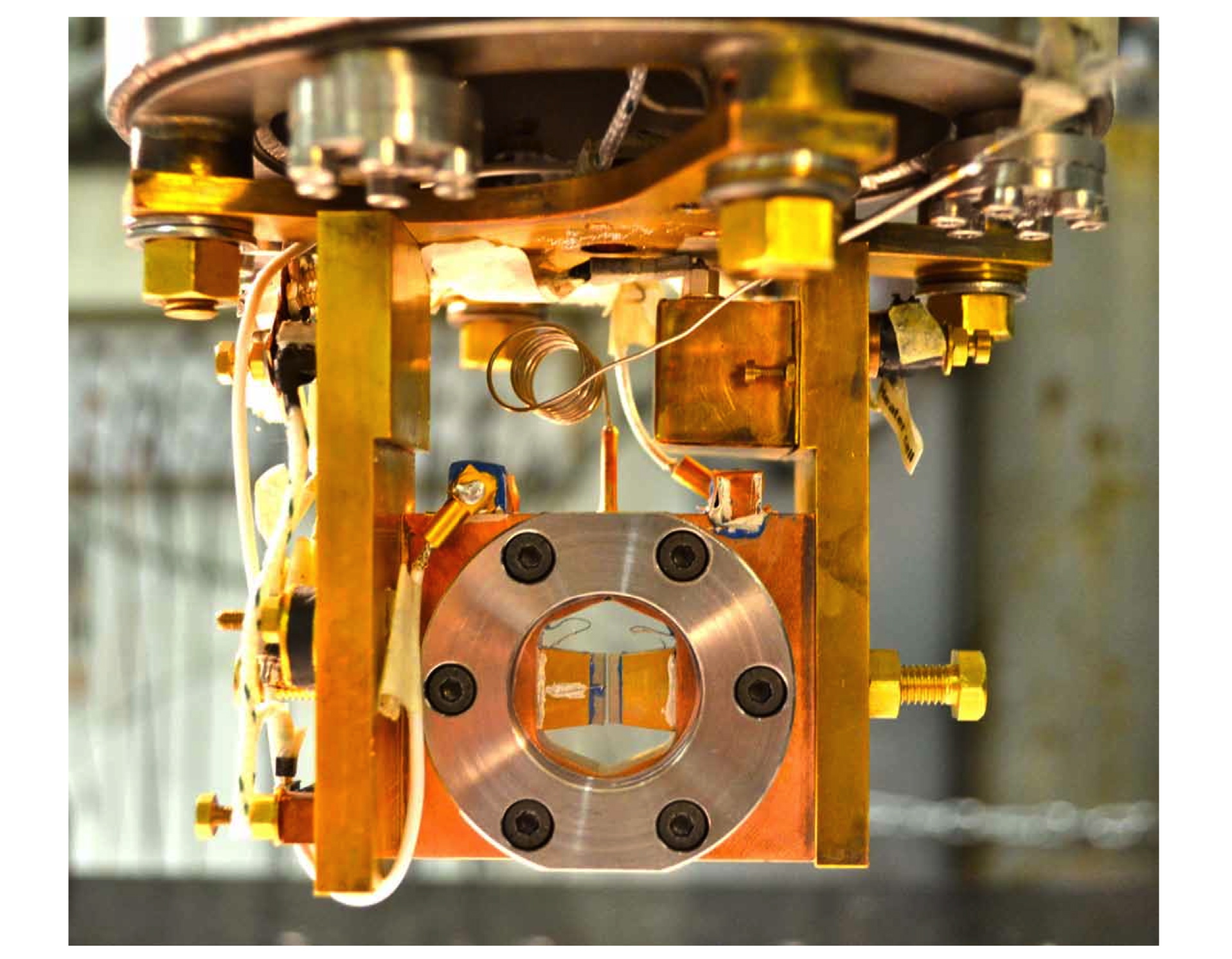}
\caption{The second cell used for this experiment is a 5 cm$^3$ hole
in a 15 mm thick Copper plate that is closed with two glass windows
sealed with stainless steel clamps and Indium rings.  The cell plate
is attached to a dilution refrigerator which allows measurements down
to 15 mK. $^4$He crystals are grown inside the 0.7 mm gap between two
piezo-electric transducers by injecting liquid through a 0.4 mm
capillary on the top (see Figure 2).  The two
transducers allow applying a vertical shear and measuring the
resulting stress across the thin crystal in the gap.  A first cell was
used for this experiment, where the gap was 1.2 mm thick.  The optical
access allows determining the crystal orientation (see Figure 3).}
\label{FigSM1-plasticity}
\end{figure}

Our experimental cell is a 5 cm$^3$ hexagonal hole in a 15 mm thick
copper plate, which is closed on its back and front faces by glass
windows sealed with Indium rings (see Figure 1) and stainless steel
clamps.  It is filled with $^4$He through a thin Cu-Ni capillary (the
"fill line" whose inner diameter is 0.4 mm).  It stands at least 62
bar inner pressure.  Inside this cell, we have glued two
piezo-electric transducers \cite{PZT} in order to shear crystals that
are grown in the gap between them.  The thickness of this gap was $d$
= 1.2 mm in a first cell used for crystals from X1 to X6 and $d$ = 0.7
mm in the second cell shown on Figure 1, where the transducers are
glued on their whole surface area in order to avoid any possible
bending near their edges.  Figure 2 shows the "loading curve", that is
the stress measured as crystallization proceeds from the bottom to the
top of the transducers.  Note that this curve shows no particular
singularity at half loading when the liquid-solid interface passes
over the two soldering points of grounding leads that are visible on
Figure 3.  One of these soldering points is on the front and the other
on the back so that there is no significant variation in the gap
thickness there.  Figure 3a shows the shape of a seed during fast
growth, which is used to determine the crystal orientation before the
crystal is regrown more slowly over the entire cell including the gap
between the two transducers (Figure 3b).  An ac-voltage (1 mV to 1V at
a frequency in the range 1 to 20 000 Hz) is applied to one transducer,
which produces a vertical displacement $u$, consequently a strain
$\varepsilon= u/d$ and finally a stress $\sigma=\mu\varepsilon$ on the
other transducer ($\mu$ is the shear modulus of the He crystal).  The
displacement $u$ is very small - of order 1 Angstrom per Volt - and it
needs to be accurately calibrated in order to obtain an absolute value
for $\mu$.  The stress generates charges, which are collected as a
current whose amplitude and phase are measured with a lock-in
amplifier.  We also use current pre-amplifiers (femto-lca-20k-200m and
femto-lca-200-10g) as was done by Day and Beamish who introduced this
method in their original work \cite{Day07}.  We improved their method
by calibrating the transducers' response in the following way.  We
first measured the cross talk between transducers.  Although the
transducer sides facing each other are grounded, there is a small
cross talk between them, which mainly comes from capacitive coupling
and needs to be accurately known as a function of frequency.  It is
independent of pressure and temperature in our working conditions so
that we measured it with the cell full of liquid.  As explained in the
main text, we then used a particular crystal (X3), which was oriented
with a [0001] axis tilted by an angle very close to 45$^\circ$ from
vertical.  For this crystal, the response to a vertical shear depends
mainly on the three elastic coefficients $c_{11}$, $c_{13}$ and
$c_{33}$, with a negligible contribution from $c_{44}$ and $c_{66}$.
The coefficients $c_{44}$ and $c_{66}$ contribute to the shear modulus
in all directions except for a shear at 45$^\circ$ from the [0001]
axis.  We then verified that the measured shear modulus was
independent of temperature in our geometry for this particular crystal
(see Fig.  3 in the main text).  Finally, we used the known values
\cite{Crepeau, Greywall} of $c_{11}$, $c_{13}$ and $c_{33}$ to obtain
the piezo-electric coefficient we needed.  In the first cell it was
$d_{15}$ = 0.88 Angstrom/Volt independently of $T$ up to 1 K and 0.95
in the second cell, about 5 times less than at room temperature.  The
shear modulus is $\mu=Id/(\omega d_{15}^2 A\sigma)$ , where $V$ is the
voltage applied to the first transducer, $A$ is the transducers' area
(1.2 cm$^2$ in the first cell and 1 cm$^2$ in the second one), $I$ is
the current generated by the stress on the second transducer.  In
order to know $I$, a careful calibration of the gain of our amplifiers
needed to be done as a function of frequency.  In the end we obtained
the absolute amplitude of the real and imaginary part of the response,
that is the shear modulus $\mu$ and the dissipation $1/Q = tan(\phi)$
where $\phi$ is the phase delay of the response.  For each crystal
orientation, we calculated $\mu$ as a function of the 5 elastic
coefficients (see below), so that we could extract the variation of
c$_{44}$.

\begin{figure}[t]
\includegraphics[width=\linewidth]{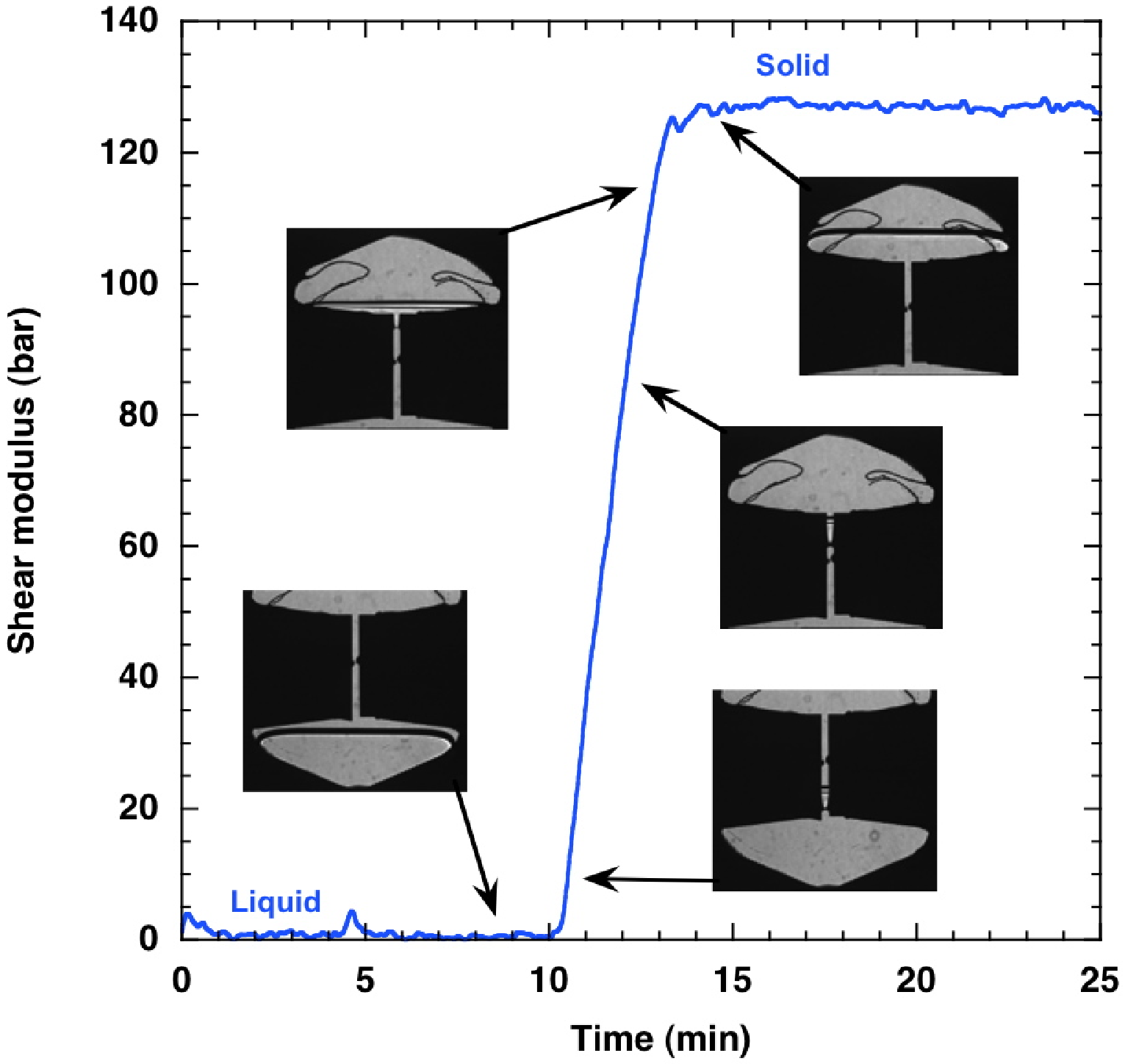}
\caption{The loading curve is the shear stress measured in the
cell as a function of time during the growth of the crystal in the
gap, which takes about 3 minutes in this case.  Near the mid height of
the transducers, one sees two point contacts for the grounding of the
transducer surfaces that face each other.  One of these contacts is on
the front and the other on the back side of the transducer, so that
there is no significant variation in the gap thickness there.  Thanks
to this design, the loading curve shows a linear variation with height
with no singularity at half loading.  }
\label{FigSM2-plasticity}
\end{figure}

\section{Sample preparation}
The quality and the purity of samples are very important in this
experiment.  The crystal quality depends on growth conditions as
previously explained by Sasaki \textit{et al.} \cite{Sasaki08} and by
Pantalei \textit{et al.} \cite{Pantalei}.  The best crystals, called
"type 1", are grown relatively slowly (up to 50 $\mu$m/s in this
experiment, 0.3 $\mu$m/s in the experiment by Rojas \textit{et al.}
\cite{Rojas}) at low $T$, usually around 20 mK, by pressurizing
superfluid $^4$He up to the liquid-solid equilibrium pressure $P_{eq}$
= 25.3 bar.  After nucleation on a random site, a crystal seed grows,
falls down to the bottom part of the cell and the growth proceeds at
constant $T$ and $P$ thanks to the mass injection into the cell
through the fill line where helium remains liquid.  This fill line is
thermally anchored along its path to the cell, so that growth does not
warm up the cell even at temperatures less than 20 mK in this
experiment.  At the equilibrium or during slow growth, the crystal
occupies the lower part of the cell with a horizontal surface and some
capillary effects where it touches walls, as would a non-wetting
liquid in a little glass bottle.  This is because the growth dynamics
proceeds with negligible dissipation and because the temperature is
highly homogeneous so that there are no temperature gradients, only a
gravity field \cite{Berghezan}.  In order to fill the cell with solid
as much as possible, one has to place the orifice of the fill line at
the highest point in the cell.  For this purpose, our cells are
tilted.  One also has to avoid corners or slits where liquid would be
trapped because of capillary effects.  Finally, it is also important
to avoid the presence of dust particles on walls because they are
efficient pinning sites for the liquid-solid interface moving up.  If
one stays at $P_{eq}$ at the end of the growth, there necessarily
remains some liquid in corners or slits, in our case at the junction
between the glass windows and the cell body.  As a consequence, the
$^3$He impurities may be trapped in this liquid if growth takes place
at low $T$, because their solubility in the liquid is much higher than
in the solid \cite{Pantalei}.  According to our experience, it is also
possible to expel all $^3$He impurities in this adjacent liquid by
applying a large ac-stress on the crystal at low $T$, because the
resulting force shakes the dislocation and detaches them from the
$^3$He atoms that are known to travel ballistically through the
crystal lattice so that they reach liquid regions in a short time and
stay trapped there.

We use a different growth procedure for "type 2" crystals.  We grow them again
at constant $T$ and $P$ from the superfluid liquid, but at 1.4 K. When the
cell is as full of solid as possible, we block the fill line by
increasing the pressure outside and we cool down to 1 K at constant
volume.  Due to the decrease of $P_{eq}$ from 26.1 bar at 1.4 K to 25.3 bar
at 1 K, the rest of the liquid in the cell crystallizes but the quality
of the final crystal, which occupies the whole cell now, is expected
to be damaged by stresses.  The advantage of this method is that it
produces a crystal in which the $^3$He concentration is known, equal to
the initial concentration in the gas cylinder, and stays at this value
during temperature cycles afterwards.  Since it is easy to melt or
grow crystals by manipulating valves outside the refrigerator, we
could melt any "type 1" crystal to a small seed and regrow it as a "type
2" crystal or vice versa.

"Type 3" crystals are polycrystals grown at constant volume from the
normal liquid above the superfluid transition.  This is known as the
"blocked capillary" (BC) method because, when cooling starts, a solid
plug forms in the fill line near the "1K pot" of the refrigerator
after what the growth proceeds at constant volume and constant mass
(but at varying P and T of course).  In this experiment, we started
cooling down the cell around 3 K with 60 bars everywhere.  In the
cell, the crystallization started at 2.4 K and finished at 1.7 K with
a final pressure P= 30 bar (see "path A" in the phase diagram of ref.
\cite{Sasaki08}).  We chose these values to avoid crossing the
reappearance of liquid near the hcp-bcc transition of the solid (see
\cite{Sasaki08}).  Indeed, recrystallization from the superfluid
usually ends up with a few large single crystals while our goal was to
obtain isotropic polycrystals with small grains and a strong disorder.
The polycrystalline nature of the sample is probably a consequence of
multiple nucleations of seeds in a cell that is far from homogeneous
in temperature in the absence of superfluid.  We found that the
crystals grown at the lowest temperature have the largest softening,
that is the largest reduction of $c_{44}$ in the soft state.  This is
probably because their dislocation density $\Lambda$ is smaller, with
a larger free length $L$ between pinning sites (a larger "pinning
length").  The precise measurement of $\Lambda$ and $L$ is in progress
in our laboratory.

\begin{figure}[t]
\includegraphics[width=\linewidth]{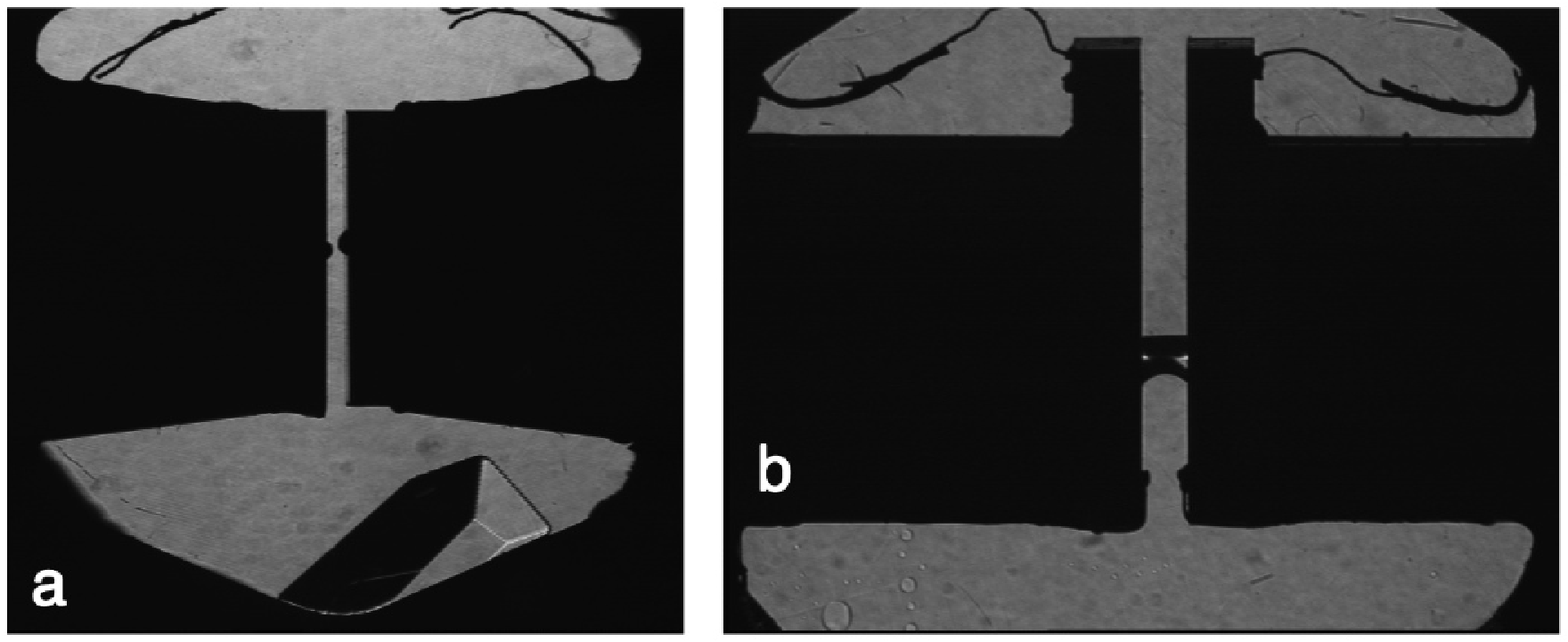}
\caption{Two photographs showing the
growth shape of a seed crystal on the left (a), and the growth inside
the gap on the right (b).  The total width of the cell is 20 mm.  The
growth of the seed is fast enough (1 mm/s) to make facets visible on
the left photograph (a) and to allow the determination of the crystal
orientation (here crystal X3 with its c axis tilted at 45$^\circ$ from
the vertical).  The growth inside the gap is shown on the right
photograph (b).  Here it is our first cell where the gap is slightly
larger, 1.2 mm instead of 0.7.  The growth inside the gap is slower
(50 $\mu$m/s).  In this first cell, the two transducers were not glued on
their entire surface and the electrical contacts were made on the
bottom edge of the transducers.  Between the two transducers, one can
see a straight horizontal line, which is the solid-liquid interface
outside the gap, and a convex one below, which shows the capillary
depression of the meniscus inside the gap.  This depression is a
consequence of a 45$^\circ$ contact angle with the transducer walls
(for more details on this partial wetting, see \cite{Balibar05}).  }
\label{FigSM3-plasticity.eps}
\end{figure}

In this experiment, we reached 80\% reduction of $c_{44}$ with the
"type 1" crystal X4 .  In a previous experiment, Rojas \textit{et al.}
\cite{Rojas} had found an 86\% reduction of $c_{44}$, after assuming
that no other elastic constant varied, for another "type 1" crystal
which was probably of even better quality for two reasons.  First, the
growth rate used by Rojas \textit{et al.} was significantly lower (0.3
$\mu$m/s) than in this experiment (50 $\mu$m/s).  Secondly, Rojas'
cell had a much more open geometry with no corners and a larger
horizontal cross section area, which allowed much more continuous
growth without sharp jumps each time the liquid-solid interface
detaches from some pinning site.

In their experiment, Ruutu \textit{et al.} \cite{Ruutu} obtained small
crystals with no screw dislocations according to their measurements of
growth rates.  Their crystals were grown slowly at 20 mK, but the
dislocation density probably depends on the growth speed just after
nucleation, which is difficult to monitor.  We have not yet succeeded
in preparing crystals without any dislocation but it is obviously an
exciting challenge because we expect the disappearance of plasticity
in that case.

\section{Orientation dependence}

A little geometry has been necessary for the data analysis. The elastic tensor of an hcp crystal involves 5 independent coefficients $c_{ij}$ with values of i,j from 1 to 6 and writes:

\begin{center}
$\begin{pmatrix}
c_{11} & c_{12} & c_{13} &0&0&0 \\
c_{12} & c_{11} & c_{13} &0&0&0 \\
c_{13} & c_{13} & c_{33} &0&0&0 \\
0&0&0& c_{44} &0&0 \\
0&0&0&0& c_{44} &0 \\
0&0&0&0&0& c_{66}
\end{pmatrix}$
\end{center}

The meaning of these indices from 1 to 6 is respectively xx, yy, zz,
yz, xz and xy with the z-axis parallel to [0001], the six-fold
symmetry axis - also called "c" - of the hexagonal structure.  The
coefficient $c_{66}$= ($c_{11}$-$c_{12}$)/2.  The orientation of the
x-axis in the plane perpendicular to the c axis is arbitrary since we
assume that the dislocations are distributed such that the transverse
isotropy of the hcp crystal is preserved.  In our experiment, the axis
c is tilted with respect to the vertical direction z' of the shear.
The crystal orientation is given by the angles $\theta$ and $\phi$ as
defined on Figure 4 where the growth shape is compared with a
hexagonal prism.  Coordinate transformations can be applied to
quantities expressed in abbreviated notation using the Bond matrices
\cite{Auld}.  Rotations of the coordinate system about its y-axis are
applied with:

\begin{center}
$M_y(\eta)=
\begin{pmatrix}
cos^2 \eta & 0 & sin^2 \eta & 0 & -sin 2\eta & 0 \\
0 & 1 & 0 & 0 & 0 & 0 \\
sin^2 \eta & 0 & cos^2 \eta &0&sin 2\eta&0 \\
0 & 0 & 0 & cos\eta &0&sin\eta \\
\dfrac{1}{2}sin 2\eta& 0 & -\dfrac{1}{2}sin 2\eta & 0 & cos2\eta &0 \\
0 & 0 & 0 & -sin\eta &0 & cos\eta
\end{pmatrix}$
\end{center}

and rotations about its z-axis are applied with:

\begin{center}
$M_z(\xi)=
\begin{pmatrix}
cos^2 \xi & sin^2 \xi & 0 & 0 & 0 & sin2\xi \\
sin^2 \xi & cos^2 \xi & 0 & 0 & 0 & -sin 2\xi \\
0 & 0 & 1 & 0 & 0 & 0 \\
0 & 0 & 0 & cos\xi & -sin\xi & 0 \\
0 & 0 & 0 & sin\xi & cos\xi & 0 \\
-\dfrac{1}{2}sin 2\xi & \dfrac{1}{2}sin 2\xi & 0 & 0 & 0 & cos2\xi
\end{pmatrix}$
\end{center}

The elastic tensor in the transducer coordinate system x'y'z' (Figure
4) is given by:

\begin{center}
$ C'=M_z(-\phi)M_y(-\theta)CM_y^T(-\theta)M_z^T(-\phi)$
\end{center}

where $X^T$ is the transpose of matrix $X$.

The shear modulus that relates the shear strain we apply to the
component of the shear stress that we measure is given by:

\begin{center}
$\mu=c'_{44}= \dfrac{1}{4} (c_{11}-2c_{13}+c_{33}) sin^2 2 \theta sin^2 \phi+c_{44}(cos^2 \theta cos^2 \phi+cos^2 2\theta sin^2 \phi)+c_{66} cos^2 \phi sin^2 \theta $
\end{center}

We used the above equation to calculate the shear modulus $\mu$ as a function of the coefficients cij for all our crystals. For example we obtained:
\begin{description}
\item $\mu=0.0001(c_{11}-2c_{13}+c_{33})+0.933 c_{44}+0.067 c_{66}$  for X2 ($\theta$ = 89.5$^\circ$ and $\phi$ = 85$^\circ$)
\item $\mu=0.25(c_{11}-2c_{13}+c_{33})+0.004 c_{44}+0.004 c_{66}$  for X3 ($\theta$ = 45$^\circ$ and $\phi$ = 85$^\circ$)
\item $\mu=0.0001(c_{11}-2c_{13}+c_{33})+0.25 c_{44}+0.56 c_{66}$  for X5 ($\theta$ = 60$^\circ$ and $\phi$ = 30$^\circ$)
\item $\mu=0.008(c_{11}-2c_{13}+c_{33})+0.97 c_{44}$  for X21  \\
($\theta$ = 5$^\circ$ and $\phi$ = 90$^\circ$)
\end{description}

The values of all c$_{ij}$ have been obtained from ultrasound velocity
measurements at 10 MHz by Crepeau \textit{et al.} at 1.32K
\cite{Crepeau} and by Greywall at 1.2K \cite{Greywall} $c_{11}$ = 405
bar, $c_{12}$ = 213 bar, $c_{13}$ = 105 bar, $c_{33}$ = 554 bar,
$c_{44}$ = 124 bar, and $c_{66}$ = 96 bar.  At such high temperatures,
the damping of dislocation motion by thermal phonons being
proportional to the frequency and to $T^3$, \cite{Day09} dislocations
cannot move at 10 MHz.  As a consequence, their values correspond to
the true elasticity of the lattice, without any contribution from
plasticity.  We have verified that, at low temperature in the presence
of $^3$He impurities, the shear modulus of our $^4$He crystals is the
same as measured by Crepeau \textit{et al.} \cite{Crepeau} and by
Greywall \cite{Greywall}.  The only correction to be made for
polycrystals grown at high pressure is the pressure dependence
analysed by H.J. Maris \cite{Maris10}.

The resolved stress $\sigma_r$ is the quantity which determines the
force acting on dislocations.  We determined it as follows.  The
strain in the transducer coordinate system (Figure 4) is given by:

\begin{center}
$E'=\begin{pmatrix}
0 & 0 & 0 & \varepsilon & 0 & 0
\end{pmatrix} ^T$
\end{center}

In the crystal frame this becomes:

\begin{center}
$E=M_y(\theta)M_z(\phi)E' $
\end{center}

The stress in the crystal frame is then :

\begin{center}
$\Sigma =CE$
\end{center}

\begin{figure}[t]
\includegraphics[width=\linewidth]{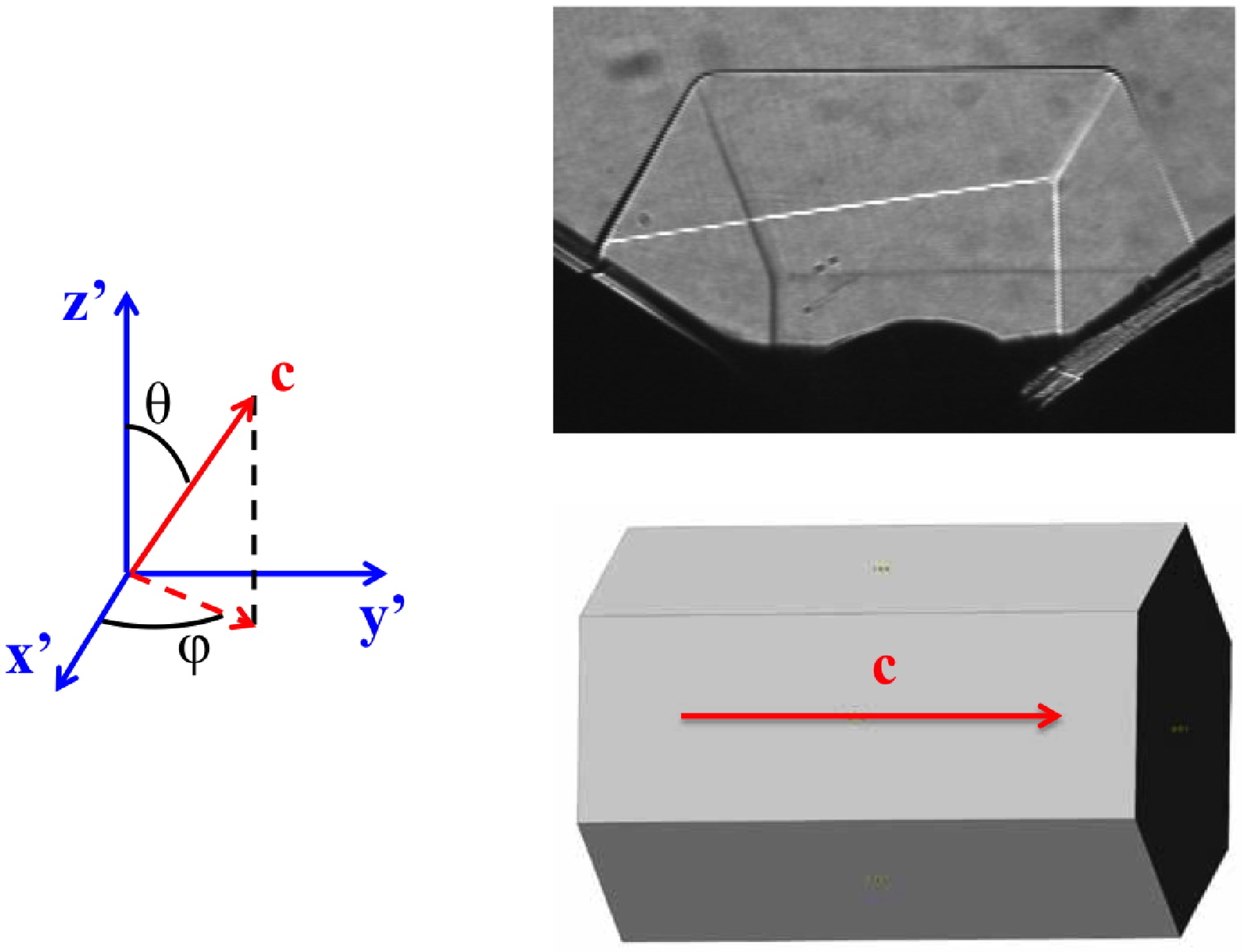}
\caption{The crystal orientation.  The orientation of the [0001] axis
"c" is defined by two angles $\theta$ and $\phi$.  The
axis z' is vertical, parallel to the transducer plane surfaces.  The
axis x' is perpendicular to the windows, and the axis y' is
perpendicular to the transducer surfaces.  The values of $\theta$ and $\phi$ are
obtained by matching the growth shape with a hexagonal prism.  In this
particular case (crystal X2), the c axis is very close to horizontal.
On the photograph, one sees that the crystal touches the front window
so that two white lines correspond to intersections of the crystal
with this front window.  Only the free edges of the crystal are used
to determine the orientation.  }
\label{FigSM4-plasticity.eps}
\end{figure}

The magnitude of the resultant of the shear stresses acting in the
basal plane $(\Sigma_4^2+\Sigma_5^2)^{1/2}$ is independent of
rotations of the crystal about its c-axis.  If we assume that the
three $\langle11\overline{2}0\rangle$ Burgers vectors are uniformly
populated then we can choose the convenient orientation with one set
of Burgers vectors along the stress resultant and the other two at
60$^\circ$ to it.  Then the average resolved stress will be:

\begin{center}
$\sigma_r=\dfrac{c_{44}\varepsilon \sqrt{cos^2\theta
cos^2\phi+cos^22\theta sin^2\phi}}{\sqrt{3}}$
\end{center}

\section{Temperature cycles and annealing of samples}
Most of the data that are analysed in this article have been obtained
by cooling samples slowly (12 hours) from 1 K down to 15 mK. The first
reason for proceeding this way is that we confirmed that some disorder
induced by mechanical perturbations at low T is annealed when warming
above 0.5 K, as was previously noticed by Day \textit{et al.}
\cite{Day09}.  Each temperature cycle took 6 hours for warming up to 1
K and 12 hours for cooling down to 15 mK, that is about one day.  We
used the same procedure for all crystals in order to include a
recording during the night when perturbations from the environment
were as small as possible (remember that we measure stresses down to 1
nanobar; He crystals are extremely sensitive to vibrations).  We then
transferred liquid Nitrogen every morning when the crystal was cold.
Every three days, a liquid helium transfer was also necessary to keep
the refrigerator working.  All these transfers produce mechanical
vibrations that shake dislocations.  After any large mechanical
perturbation at low temperature, we observe some hardening of
crystals, which we believe is due to creation of jogs on the
dislocations.  After annealing up to 1 K we found reproducible results
as if jogs had been eliminated thanks to the diffusion of thermally
activated vacancies.

\section{Comparison with a classical crystal}

In order to compare the plasticity in helium crystals and in classical
crystals, let us consider the historical measurements by Tinder and
Washburn\cite{Tinder} on Copper.  Qualitatively, the plasticity is the
same phenomenon: it is due to the motion of dislocations and it
is sensitive to the concentration of impurities.  But quantitatively, there are striking differences.
Tinder and Washburn found a threshold stress of about 2~g/mm$^2$ = 2
10$^4$ Pa beyond which a plastic strain appears in addition to the
usual elastic response.  This threshold is somewhat smaller than in
other classical crystals, probably because of the careful growth and
manipulation of these very pure samples.  Still, it is larger by five
orders of magnitude than in our case. In Copper, the applied stress
is 0.4~10$^{-6}$ times the
shear modulus $\mu_{Cu}$ = 50~GPa.  For helium crystals, we find a
linear response (no threshold) down to 1~nanobar = 10$^{-4}$~Pa, which is
10$^{-11}$ times the elastic shear modulus $\mu_{He}$~=~12~MPa.
Furthermore, Tinder and Washburn find a plastic strain that is 40
times \textit{smaller} than the elastic strain so that the effective
shear modulus is not significantly changed.  In Helium, the plastic
strain due to the dislocation motion is 4 times \textit{larger} than
the elastic strain, leading to an effective shear modulus that is
reduced by 80\%. In other words, the plastic response is 2 orders of
magnitude larger for stresses 5 orders of magnitude smaller than in
Copper.  In Copper, the plastic response is highly non-linear and the
response time of order minutes at room temperature (300~K).
In Helium at 0.1~K, the plasticity is linear
so that it results in an effective reduction of the shear modulus,
which we have found independent of frequency up to 16~kHz.
We have been able to study oriented
single crystals and we have found evidence that the gliding plane of
dislocations is the basal plane, so that the plasticity is
anisotropic. Tinder and Washburn studied polycrystalline samples where
they could not measure any orientation dependence of the plasticity.

This work was supported by grants from ERC (AdG
247258-SUPERSOLID) and from NSERC Canada.